\documentstyle[preprint,eqsecnum,aps,epsf]{revtex}	

\newif\iftightenlines\tightenlinesfalse
\tightenlines\tightenlinestrue

\begin{document}
%
\def\eslt{E\llap/_T}
\def\esl{E\llap/}
\def\slashB{B\llap/}
\def\msl{m\llap/}
\def\to{\rightarrow}
\def\te{\tilde e}
\def\tmu{\tilde\mu}
\def\ttau{\tilde\tau}
\def\tl{\tilde\ell}
\def\ttau{\tilde \tau}
\def\tg{\tilde g}
\def\tnu{\tilde\nu}
\def\tell{\tilde\ell}
\def\tq{\tilde q}
\def\tb{\tilde b}
\def\tst{\tilde t}
\def\tt{\tilde t}
\def\tw{\widetilde W}
\def\tz{\widetilde Z}
\def\beq{\begin{equation}}
\def\eeq{\end{equation}}
\def\beqa{\begin{eqnarray}}
\def\eeqa{\end{eqnarray}}
\newcounter{temp}
\def\baeq{\setcounter{temp}{\theequation}
\addtocounter{temp}{1}
\setcounter{equation}{0}
\renewcommand{\theequation}{\arabic{temp}\alph{equation}}}
\def\eaeq{\setcounter{equation}{\thetemp}
\renewcommand{\theequation}{\arabic{equation}}}

\hyphenation{mssm}
%

\preprint{\vbox{\baselineskip=14pt%
   \rightline{UH-511-888-97}
   \rightline{APCTP 97-19}
}}
\title{ A Supersymmetric Resolution of Solar and Atmospheric
Neutrino Puzzles}
\author{ M. Drees$^1$, S. Pakvasa$^2$, X. Tata$^2$ and T. ter Veldhuis$^2$}
\address{$^1$APCTP, 207--43 Cheongryangri--dong, Seoul 130--012, Korea}
\address{
$^2$Department of Physics and Astronomy,
University of Hawaii,
Honolulu, HI 96822 USA
}

%
\maketitle
\begin{abstract}

Renormalizable lepton number violating interactions that break
R--parity can induce a Majorana mass for
neutrinos. Based on this, we show that it is possible to
obtain a phenomenologically viable neutrino mass matrix
that can accommodate atmospheric neutrino data via 
$\nu_{\mu}$--$\nu_{\tau}$ mixing
and the solar neutrino data via either the large
or small angle MSW effect. We argue that such a mass matrix could result
from an approximate
discrete symmetry of the superpotential that
forbids renormalizable baryon number violating couplings.
\end{abstract}
\newpage
\section{Introduction and Phenomenology}
The deficit of solar neutrinos\cite{solar},
the observed ratio of $\nu_{\mu}$ to $\nu_{e}$ events from 
atmospheric neutrinos\cite{atmos}, and the signal from the
LSND experiment\cite{LSND} all point toward non--trivial mixing effects
amongst neutrinos which, in turn, implies
that not all neutrinos can be massless.
The flux of solar neutrinos may be accounted for by MSW oscillations
provided\cite{HL} either 
$\Delta m^2 \sim 2\times 10^{-5}$~eV$^2$ and the mixing angle is large,
or $\Delta m^2 \sim (7 \pm 3)\times 10^{-6}$~eV$^2$ with $\sin^2 2 \theta
\sim 7\times 10^{-3}$. The explanation of the atmospheric neutrino puzzle,
on the other hand, appears to require $\Delta m^2 \sim 5\times 10^{-3}$~eV$^2$
with large mixing between $\nu_{\mu}$ and another neutrino. 
Recently, the CHOOZ\cite{CHOOZ} collaboration has excluded
the mixing of $\nu_e$ with another neutrino if $\Delta m^2 > 2\times 10^{-3}$~eV$^2$
and $\sin^2 2 \theta > 0.2$. Thus $\nu_{\mu}$--$\nu_{\tau}$
is the favoured explanation of the Super Kamiokande data\cite{Tot}.
Finally the LSND data\cite{LSND} require a much larger mass difference 
$\Delta m^2 \sim 1 $~eV$^2$ (and small mixing). 
The LSND effect will, in the near future, be
independently probed by the KARMEN experiment\cite{KARM}.

Three distinct mass differences can only
be accommodated in models with at least four neutrino flavours
(one of them sterile), except when two of the phenomena
\lq\lq share\rq\rq the same mass difference~\cite{CF,AP}, although this
interpretation is now disfavoured. We take a much less ambitious 
approach in this paper, and exclude the LSND observation
from our consideration. We point out that
supersymmetry (SUSY) models in which
R--parity is explicitly broken by
lepton number violating superpotential interactions offer
a novel approach for the
construction of phenomenologically viable neutrino
mass matrices. We show that it is possible to
obtain a neutrino mass matrix that is in accord with solar
neutrino and atmospheric neutrino data, without any need to
assume a hierarchy between different R--parity
violating couplings. To achieve this, we are led
to impose additional global (discrete)
symmetries that also automatically forbid dimension
four baryon number violating (\slashB) interactions.

Assuming the field content of the minimal supersymmetric model, 
the most general R--parity violating 
trilinear superpotential can be written as,
\beq
f= f_1 + f_2 + f_3, 
\eeq
with
\beq
f_1 = \lambda_{ijk} L_i L_j E_k^C, \label{f1}
\eeq
\beq
f_2 = \lambda_{ijk}' L_i Q_j D_k^C, \label{f2}
\eeq
and
\beq
f_3 = \lambda_{ijk}'' U_i^C D_j^C D_k^C. \label{f3}
\eeq
The superpotential interactions (\ref{f1}) and (\ref{f2})
provide us the desired source of lepton number violation
while the simultaneous presence of those
in (\ref{f3}) is dangerous since they can lead to proton
decay at the weak rate. As mentioned above,
we will show that the couplings in (\ref{f3}) can
be forbidden as a result of a symmetry.

We begin our analysis with the observation~\cite{DH,GRT}
that $f_1$ and $f_2$ superpotential interactions
lead to Majorana masses for neutrinos. For
example, the $\lambda'$ couplings lead to a Majorana
mass
\beq
m_{\nu_{i}} \sim \frac{3}{8 \pi^2} \lambda_{ijk}' \lambda_{ikj}'
\frac{1}{m_{\tilde{q}}^2} M_{SUSY} m_j m_k
\eeq
for the $i^{th}$ neutrino via diagrams involving a 
quark--squark loop. Here, $i=e,\mu$ or $\tau$,
$m_j$ ($m_k$) is the mass of the down--quarks
in the $j^{th}$ ($k^{th}$) generation and 
$M_{SUSY}$ ($\sim A \sim \mu$, where $A$ and $\mu$ are the
usual SUSY parameters) is a mass--scale
that determines mixing between $\tilde{q}_L$ and $\tilde{q}_R$.
Notice that (in the absence of a miraculous cancellation)
a non--zero mass for $\nu_i$ results if $\lambda_{ijj}' \neq 0$,
whereas for $j\neq k$, $\lambda'$ interactions induce a mass
only~\cite{BC} when $\lambda_{ijk}' \lambda_{ikj}' \neq 0$. We also remark
that the $\lambda$ interactions can similarly induce neutrino
masses.

The R--parity violating couplings in $f_1$ and $f_2$ can
also induce off--diagonal Majorana neutrino mass terms
which violate both lepton number and flavour conservation.
This is the origin of the neutrino mass matrix
that we will analyse below.

The large number (9+27) of {\it a priori} unconstrained
$\lambda$ and $\lambda'$ \  couplings makes it apparent that there
is a lot of freedom in the neutrino mass matrix. We will,
however, adopt the philosophy that 
all the {\it new} superpotential couplings that are allowed
by the symmetries that we impose should be
comparable in magnitude; {\it i.e.} there is no large hierarchy
between these couplings. The dominant
contribution to neutrino masses then comes from $b \tilde{b}$ loops,
and the neutrino mass matrix takes the form,
\beq
m_{\nu} \sim \frac{3}{8 \pi^2} \frac{m_b^2}{m_{\tilde{q}}^2}
M_{SUSY}
\left(
\begin{array}{ccc}
\alpha_e^2            & \alpha_e \alpha_{\mu} & \alpha_e \alpha_{\tau} \\
\alpha_e \alpha_{\mu} & \alpha_{\mu}^2     & \alpha_{\mu} \alpha_{\tau} \\
\alpha_e \alpha_{\tau} & \alpha_{\mu} \alpha_{\tau} & \alpha_{\tau}^2  
\end{array}
\right), \label{massmat}
\eeq
where we have introduced the short--hand notation 
$\lambda_{i33}' \equiv \alpha_i$.
The mass matrix in (\ref{massmat}) is valid
up to corrections which are suppressed by
$m_s/m_b$ (from $\lambda_{i23}'$ and $\lambda_{i32}'$
couplings) and comparable corrections $\sim m_{\tau}^2/3 m_b^2$
(from $\lambda_{i33}$ couplings).

The interesting property of the matrix in (\ref{massmat}) is that
regardless of the precise values of $\alpha_i$, two of its eigenvalues
are zero. Corrections of ${\cal{O}}(\frac{m_s}{m_b} \sim 
\frac{m_{\tau}^2}{3 m_b^2})$, in general, will
cause these eigenvalues to shift from zero,
and we may thus expect their mass difference
(which should be comparable in magnitude to the
individual masses) to have a similar ratio
to the large mass difference. For $m_s = 200$~MeV
and $m_b = 5$~GeV this is in remarkable agreement with
the ratio $\Delta m_{solar}/\Delta m_{atmos}$ that we
discussed in the beginning of this paper.

This cannot, however, be the complete story. In order to accommodate
the Super Kamiokande data\cite{Tot}, the mixing of $\nu_e$ should be
small. Then, to account for the $50\%$ change in the
``ratio of ratios'' from SM expectation,
we are required to have $\nu_{\mu}$--$\nu_{\tau}$ mixing close
to its maximal value. We are, therefore, led to the 
phenomenological constraint,
\begin{displaymath}
(m_{\nu})_{12},  (m_{\nu})_{13} << (m_{\nu})_{22} \simeq (m_{\nu})_{33}
\simeq (m_{\nu})_{23}. \nonumber
\end{displaymath}
Since our philosophy does not allow us to include 
a hierarchy between the various $\lambda'$ couplings,
we will seek models where $\lambda_{133}'=0$ due
to a symmetry that allows $\lambda_{233}'$ and $\lambda_{333}'$.
In this case the magnitudes of $(m_{\nu})_{12}$ and $(m_{\nu})_{13}$
will be set by $m_b m_s$ (provided the appropriate
$\lambda'$s do not vanish) and we will obtain
a neutrino mass matrix of the form,
\beq
m_{\nu} \sim \frac{3}{8 \pi^2} \lambda' \cdot \lambda'
\frac{m_b^2}{m_{\tilde{q}}^2}
M_{SUSY}
\left(
\begin{array}{ccc}
\frac{m_s^2}{m_b^2}   & \frac{m_s}{m_b}  & \frac{m_s}{m_b}  \\
\frac{m_s}{m_b}  & 1     & 1 \\
\frac{m_s}{m_b}  & 1     & 1  
\end{array}
\right), \label{massmat2}
\eeq
which, as we have already discussed, readily accounts for the
data provided
\begin{displaymath}
\frac{3}{8 \pi^2} \lambda' \cdot \lambda' 
\frac{m_b^2}{m_{\tilde{q}}^2} M_{SUSY} \sim
(5\times 10^{-3})^{1/2} / {\rm eV}. \nonumber
\end{displaymath}
Taking $m_{\tilde{q}}=M_{SUSY}$, 
and assuming all the non--vanishing $\lambda'$ couplings have
comparable magnitude, we have
\beq
{\lambda'} \sim 10^{-4} (\frac{m_{\tilde{q}}}{200 \ {\rm GeV}})^{1/2}.
\eeq
It should be recognized that the mass matrix in (\ref{massmat2}) may
be modified if the $\lambda$ couplings in (\ref{f1}) are
also non--vanishing, since these interactions can induce
a mass 
$\sim \frac{1}{8 \pi^2} \lambda \cdot \lambda
\frac{m_{l_i} m_{l_j}}{m_{\tilde{l}}^2} M_{SUSY}$.
But as long as $m_{\tilde{l}} \simeq m_{\tilde{q}}$ and
$\lambda \alt \lambda'$, the qualitative features of the neutrino
mass matrix (\ref{massmat2}) are not altered since
$3 m_s/m_b \simeq m_{\tau}^2/m_b^2$.

Before turning to a discussion of symmetries of
the superpotential that could yield a neutrino 
mass matrix with the form in (\ref{massmat2}), we briefly
discuss other aspects of the phenomenology of the 
model:
\begin{enumerate}
\item Neutrino masses are too tiny to allow for their 
direct detection.
\item R--parity violating couplings $\lambda$ or $\lambda' 
\sim 10^{-4}$ will result in the lightest supersymmetric particle
(LSP), assumed to be the lightest neutralino, decaying
inside the detector~\cite{Daw}, probably without an observable displaced vertex.
We do not expect the
production or decays of SUSY particles (other than the LSP) to be modified
by these tiny couplings.
\item LSP pair production can potentially lead to observable 
signals at LEP2, even if other sparticles are
kinematically inaccessible\cite{DH}. Some bounds have already been
obtained by the ALEPH Collaboration~\cite{aleph}. The R--parity violating
interactions will also modify SUSY signals at hadron
colliders, but exactly how these alter will depend on
the decay pattern of the LSP. While the canonical 
$\eslt$ signal will be reduced relative to the usual expectation,
signals in the same sign dilepton and multilepton (plus multijet)
channels should be readily observable, both at
the Tevatron\cite{R1} and at the LHC\cite{R2}.
\item The R--parity violating interactions in (\ref{f1}) and
(\ref{f2}) induce lepton number violating 
sneutrino masses, $\delta m^2 \simeq \frac{3}{16 \pi^2} |\lambda'|^2
m_b^2$ or $\frac{1}{16 \pi^2} |\lambda|^2 m_{\tau}^2$
(the $m_b^2$ term is only possible for $\tilde{\nu}_{\mu}$ and
$\tilde{\nu}_{\tau}$, since $\lambda_{133}'=0$) that possibly also
violate flavour.
Unfortunately, since sneutrino decays are generically governed
by their gauge couplings which are much larger
than $10^{-4} \times m_b/M_W$, sneutrinos will decay much before they can
even begin to oscillate.
\item Finally, because we are allowing interactions where
more than one lepton flavour is not conserved, we
have to worry that these interactions do not 
induce decays such as $\mu \rightarrow e \gamma$ or
$K^0 \rightarrow \mu e $ whose branching ratios are
strongly constrained~\cite{pdg}.
The lowest dimensional effective operator that causes the 
decay $\mu \rightarrow e \gamma$ has the form 
$K \bar{e} \sigma_{\mu \nu} \mu F^{\mu \nu}$, and
necessarily flips the lepton chirality. If lepton number
is violated only by $\lambda'$ interactions, only left
handed leptons couple via these, so that $K$ must include a
lepton mass to flip the chirality. 
Dimensional analysis then tells us that 
$K \sim \frac{3 {\lambda'}^2}{8 \pi^2 m_{\tilde{q}}^2} m_{\mu}$,
where we have included a loop factor $\sim 8 \pi^2$
and a color factor of $3$.
The suppression factor $1/m_{\tilde{q}}^2$ in $K$ then implies
$\Gamma(\mu \rightarrow e \gamma) \sim e_q^2 \frac{\alpha}{4}
(\frac{3 {\lambda'}^2}{8 \pi ^2})^2 
\frac{m_{\mu}^5}{m_{\tilde{q}}^4}
\times {\cal{O}}(1)$, so that in order not to violate the experimental
bound~\cite{pdg} 
$B(\mu \rightarrow e \gamma) < 5 \times 10^{-11}$, we must have
${\lambda'}^4 \alt 10^{-9}$ for 
$m_{\tilde{q}}=200$~GeV. We thus see that $\lambda'$ couplings
$\sim 10^{-4}$ as required by the neutrino data are
comfortably within this bound even if $\lambda_{133}'$ is comparable
in magnitude to $\lambda_{233}'$ and $\lambda_{333}'$. Notice
that there is no bound if $\lambda_{1jk}' \neq 0$ only for
those $\{ jk \}$ values for which $\lambda_{2jk}'$ vanishes, as could well
be the
case for us. Bounds on violation of $\tau$ number are weaker
than those from $\mu \rightarrow e \gamma$ decay and do not pose serious
constraints for the magnitudes of couplings relevant to us. We
will revisit this decay when we consider specific models
below and allow for the possibility of non--vanishing
$\lambda$--type couplings.
\item The superpotential $\lambda'$ couplings can
potentially also mediate $K^0 \rightarrow \mu e$ decays via
virtual squark exchange. Once again the amplitude will be proportional
to an explicit fermion mass factor. This amplitude is
non--vanishing only when both $\lambda_{12k}'$ and
$\lambda_{21k}'$ or $\lambda_{1j2}'$ and $\lambda_{2j1}'$ are
simultaneously non--vanishing. We will
return to this model--dependent issue
once we set up the framework that yields the mass
matrix of the desired form.
\end{enumerate}

\section{Models}
Our strategy is to seek a discrete symmetry of the superpotential
of the model which allows the desired Yukawa interaction of the
Standard Model as well as $\lambda$ or $\lambda'$ interactions that yield
a neutrino matrix of the form in (\ref{massmat2}); {\it i.e.} allows
couplings such as $\lambda_{233}'$ and $\lambda_{333}'$ but
forbids $\lambda_{133}'$  couplings which would lead to unit entries
instead of $m_s/m_b$ and $m_s^2/m_b^2$.
We will also require that the symmetry forbid the $\slashB$ superpotential
interactions (\ref{f3}). We will, however, see that it is {\it not
possible} to find an Abelian symmetry that does the job. We will give
general 
arguments to show that {\it any such symmetry has to be broken
by the strange quark Yukawa couplings}. For
definiteness, we will work with $Z_N$ symmetries , but
the generality of these arguments will be clear. We
write the superpotential interactions in the mass basis~\cite{fn1}
of quarks, not in the current basis.

\subsection{$Z_2$ symmetry}
In this case each superfield carries a multiplicative 
quantum number $+1$ or $-1$. We can choose the
quantum number of $L_3$ to be $+1$ by convention. We will
denote this as $L_3=+1$. Then, since 
$\lambda_{333}' \neq 0$, $Q_3 D_3^C = +1$.
It then follows that to get maximal $\nu_{\mu}$--$\bar{\nu}_{\tau}$
mixing, we must have $L_2=+1$, and $L_1=-1$. This already
shows that our $Z_2$ symmetry must be broken in order to explain solar 
neutrinos by mixing between 
$\nu_e$ and $\nu_{\mu}$ or $\nu_{\tau}$. But let us proceed further to
understand how this is broken. A non--vanishing bottom Yukawa
coupling means $h'=+1$ so that a non--vanishing $\mu$
term implies $h=+1$, where $h'$ and $h$ are the Higgs fields
whose interactions give a mass to  $T_3=-1/2$ and $T_3= 1/2$
fermions, respectively. Finally, the large top quark mass
implies $Q_3 U_3^C = +1$. Thus $Q_3$, $U_3^C$ and $D_3^C$ all have the same
charge.

Now, let us forbid the unwanted $\lambda''$ couplings. We first note
that $D_{1,2}^C=-1$ to forbid $U_3^C D_3^C D_{1,2}^C$ coupling, which
together with the absence of $U_i^C D_1^C D_2^C$ term requires
$U_i^C=-1$. Thus all the quark singlets have charge $-1$.

In order to get $(m_{\nu})_{12} \sim (m_{\nu})_{13} \propto m_b m_s$,
we must have $L_1 Q_2 D_3^C$ and $L_{2,3} Q_3 D_2^C$ interactions,
because the $L_1 Q_3 D_2^C$ term is forbidden. This forces us to choose
$Q_2=+1$, so that both $m_s$ and $m_c$ are forbidden!
Since the charm Yukawa coupling ($\sim10^{-2}$) is two
orders of magnitude larger than the $\lambda'$ couplings
that we require, this is not very appealing.

One way to ameliorate this situation is to
change the scale of the mass factor outside the matrix (\ref{massmat2})
from $m_b^2$ to $m_b m_s$ since this will increase
${\lambda'}^2$ by $m_b /m_s$. But this means that a $Z_2$ symmetry
has to forbid $L_i Q_3 D_3^C$ couplings for all values of $i$ so that
$L_1=L_2=L_3$ and $\nu_e$--$\nu_{\tau}$ and $\nu_e$--$\nu_{\mu}$ 
mixing is unsuppressed.

Continuing this line of thought it is natural to
ask whether it is possible to change the mass
scale in the matrix in (\ref{massmat2}) to $m_s^2$ and the
ratio $m_s/m_b$ to $m_d/m_s$ which coincidentally has a 
similar magnitude. This would allow a yet larger $\lambda'$
coupling, and also offer the hope that we can find a symmetry
that is only violated by $m_d$ and
not $m_s$. But this is impossible for any symmetry
in {\it any} theory where the $b$- and $s$- quarks get
their masses from the same Higgs boson because then,
$Q_2 D_2^C = Q_3 D_3^C$. Thus, if $L_{2,3} Q_2 D_2^C$ terms are
allowed by the symmetry, so are $L_{2,3} Q_3 D_3^C$ terms,
so that the scale of neutrino masses will be as in (\ref{massmat2}).

\subsection{$Z_3$ symmetry}
While the $Z_2$ symmetry that we found indeed gave
us the mass matrix (\ref{massmat2}) it suffered from the
fact that the charm quark Yukawa coupling (which is forbidden
by this symmetry) far exceeds the $\lambda'$ couplings that
are phenomenologically required and allowed by the
symmetry.
The reader can check that requiring instead a $Z_3$
symmetry with additive charges (modulo 3),
\beqa
U_i^C & = & Q_i= D_3^C=h=h'=L_2=L_3=E_3=0 \nonumber \\
D_1 & = & D_2 = 1 \nonumber \\
L_1 & = & E_1=E_2=2 \nonumber
\eeqa
\begin{enumerate}
\item top, charm, bottom and tau Yukawa couplings are allowed
\item we obtain a neutrino mass matrix whose order of magnitude 
of entries is given by (\ref{massmat2}); $(m_{\nu})_{22}$ gets
an additional contribution from the $L_2 L_3 E_3$ coupling.
Nevertheless, the situation is still problematic because
the couplings $\lambda' \sim 10^{-4}$ are still an
order of magnitude smaller than the strange
quark Yukawa coupling. We are thus forced to seek a model
where the scale of the neutrino mass matrix is $m_s^2 / m_{\tilde{q}}$
rather than $m_b^2/m_{\tilde{q}}$ as in (\ref{massmat2}).
\end{enumerate}

Toward this end, we consider a $Z_3$ symmetry with charges
given by,
\beqa
U_i^C & = & D_3^C = Q_2 = Q_3 = h=h'=0 \nonumber \\
L_3 & = & L_2=E_2=1 \nonumber \\
E_3 & = & D_2^C=D_1^C=Q_1=L_1=E_1=2 \nonumber
\eeqa
This symmetry
\begin{enumerate}
\item allows $m_t$, $m_b$, $m_c$ and $m_{\tau}$, but is explicitly broken
by other quark masses;
\item forbids all renormalizable $\slashB$ interactions;
\item forbids lepton number violating bilinears in the superpotential,
which gives us an {\it a posteriori} justification for ignoring these;
\item forbids $L_i Q_3 D_3^C$ couplings, and since 
$Q_2 D_3^C \neq Q_3 D_2^C$ also
simultaneous presence of $L_i Q_2 D_3$ and $L_i Q_3 D_2$ (which
would give $m_b m_s$ terms in the neutrino mass matrix); $L_i L_3 E_3$
couplings are also forbidden;
\item allows $L_{2,3} Q_2 D_2^C$ but forbids
$L_1 Q_2 D_2^C$ couplings in (\ref{f2});
\item allows $L_1 Q_1 D_2^C$ and $L_{2,3} Q_2 D_1^C$
couplings which allows $\nu_e$ to mix with $\nu_{\mu,\tau}$ via entries
proportional to $m_d m_s/m_{\tilde{q}}$.
\item Only the $L_2 L_3 E_2^C$ couplings in $f_1$ are allowed;
this gives a correction to the $(m_{\nu})_{33}$ entry.
\item Finally, notice that $L_1 D_2^C \neq L_2 D_1^C$ and
$L_1 Q_2 \neq L_2 Q_1$. Thus the potentially dangerous 
$K^0 \rightarrow \mu e$ decays are absent in this framework.
\end{enumerate}

The neutrino mass matrix for this model has the form,
\beq 
m_{\nu} \sim \frac{3}{8 \pi^2} \lambda' \cdot \lambda'
\frac{m_s^2}{m_{\tilde{q}}^2}
M_{SUSY}
\left(
\begin{array}{ccc}
\frac{m_d^2}{m_s^2}   & \frac{m_d}{m_s}  & \frac{m_d}{m_s}  \\
\frac{m_d}{m_s}  & 1     & 1 \\
\frac{m_d}{m_s}  & 1     & 1+\delta  
\end{array}
\right), \label{massmat3}
\eeq
where $\delta \sim \frac{\lambda \cdot \lambda}{\lambda' \cdot \lambda'}
\frac{m_{\mu}^2}{3 m_s^2}
\sim \frac{1}{10} \frac{\lambda \cdot \lambda}{\lambda' \cdot \lambda'}$
and so can easily be comparable to $m_d/m_s \sim 8 \ {\rm MeV}/200 \ {\rm MeV}
\sim 1/25$.
We also remind the reader that because the $\lambda' \cdot \lambda'$
factors that multiply each entry in the
matrix (\ref{massmat3}) need not be identical, it is the {\it form}
of the matrix and not the exact matrix elements listed
that we should focus on. In the limit that the $m_d/m_s$ and
$\delta$ terms are neglected, we have one massive neutrino
state with $m_{\nu} \sim \frac{3}{4 \pi^2} \lambda' \cdot \lambda'
\frac{m_s^2}{m_{\tilde{q}}^2} M_{SUSY}$ and two massless states.
Furthermore, for equal $\lambda'$s this massive state is a maximal
mixture of $\nu_{\mu}$ and $\nu_{\tau}$ as required by
Super Kamiokande data\cite{Tot}.
Since we now have $m_s^2$, not $m_b^2$, setting the scale of the
neutrino mass matrix, $\lambda'$ will now be larger by a factor
$\sim m_b/m_s \sim 25$ compared to our earlier estimate; {\it i.e.}
$\lambda' \sim 2.5 \times 10^{-3}$, somewhat larger than the strange quark
Yukawa coupling.
While the situation is not ideal, considering factors of ${\cal{O}}(1)$
that we have ignored, it is probably not inconsistent.

The $m_d/m_s$ and $\delta$ terms in (\ref{massmat3}) result in a mass for
the other states. The exact values of the masses are sensitive
to the precise values of the various $\lambda'$ couplings, but the
{\it mass difference} between them is suppressed (relative to the 
\lq\lq large\rq\rq mass) by a factor of ${\cal{O}}(m_d/m_s$ or $\delta)$,
and so, is exactly in the right ballpark to account for the solar neutrino
flux via
the MSW effect. The mixing between $\nu_e$ and the other
\lq\lq massless\rq\rq state due to the $m_d/m_s$ and $\delta$
terms is sensitive to model parameters but is generically large, so
in general, we may expect solar neutrino to be accounted for
by the large angle MSW solution. Its, however, not
implausible that the mixing angle becomes suppressed
due to cancellations between $(m_{\nu})_{12}$ and
$(m_{\nu})_{13}$ terms
(which are expected to have very similar magnitude) when
this mixing is estimated treating $m_d/m_s$ and $\delta$ to
first order in perturbation theory. A cancellation precise to 
$\sim 10 \%$ can then account for solar neutrinos as a small angle
MSW effect.
As is well known, the two scenarios of small and large mixing MSW
solutions
can be distinguished experimentally by the energy spectrum
distortion expected in the former, and the
day night effect in the latter, both for $^8B$ neutrinos.

The larger magnitude, $\lambda' \sim 2.5\times 10^{-3}$, of the superpotential
couplings may cause renewed concern about the experimental
bound on the decay rate for $\mu \rightarrow e \gamma$. The reader can
easily check that because of the $Z_3$ symmetry (even though
it is explicitly broken) there is no 1-loop diagram that
mediates this decay. Moreover, $\lambda$--interactions conserve
$e$ and $\mu$ number so that they cannot contribute. 
Finally, exactly as before, the oscillation length of 
sneutrinos is still larger than their decay length.

\section{Summary}
In summary, we have proposed a novel mechanism involving
R--parity violation for generating neutrino mass matrices
consistent with solar and atmospheric neutrino data. The observed
ratio of ratios of atmospheric neutrinos is due to maximal
$\nu_{\mu}$--$\nu_{\tau}$ mixing. We favour large angle MSW oscillations
as the solution to the solar neutrino puzzle but do not exclude
the small angle MSW solution. Future measurements will help discriminate
between various scenarios and pin down the model parameters.
We have argued that
the structure of the neutrino mass matrix reflects an
underlying approximate symmetry of the superpotential.
Although our mechanism does not appear to have direct
implications for low energy physics (other than the 
neutrino phenomenology already discussed), the new superpotential 
interactions will significantly alter SUSY signals
at high energy colliders.

\acknowledgements
We thank John Learned for continually pressing us to think about neutrino 
data. Two of us (SP and XT) are grateful to the Asia Pacific Centre for 
Theoretical Physics, Seoul, for hospitality while this work was in progress. This 
research was supported in part by the United States Department of Energy 
Grant DE-FG-03-94ER40833.


\begin{thebibliography}{99}
\bibitem{solar} See talks by the Homestake, Kamiokande, Sage
                and Gallex Collaborations in: 
                Proceedings XVII International Conference on 
                Neutrino Physics and Astrophysics, Helsinki Finland
                (13--19 June 1996), eds. J. Maalampi and M. Roos
                (World Scientific, Singapore, 1997), to be published.
\bibitem{atmos} K. S. Hirata {\it et al.}, Phys. Lett. {\bf B280}, 146 (1992).
                D. Casper {\it et al.} Phys. Rev. Lett. {\bf 66}, 2561 (1991).
                SOUDAN collaboration, W.W.M. Allison {\it et al.}
                Phys. Lett. {\bf B391}, 491 (1997);
                NUSEX Collaboration, M. Aglietta {\it et al.}
                Europhysics. Lett. {\bf 8}, 611 (1989);
                FREJUS Collaboration, Ch. Berger {\it et al.} Phys. Lett.
                {\bf B227}, 489 (1989) and Phys. Lett. {\bf B245}, 305 (1990).
\bibitem{LSND}  The LSND collaboration, C. Athanassapoulos {\it et al.}
                Phys. Rev. {\bf C54}, 2685 (1996) and  nucl-ex/9709006.
\bibitem{HL}    N. Hata and P. Langacker, Phys. Rev. {\bf D56}, 6107
                (1997).
\bibitem{CHOOZ} The CHOOZ Collaboration, M. Appollonio {\it et al.}
                hep-ex/9711002,  Phys. Lett. {\bf B} (in press).
\bibitem{Tot}   Y. Totsuka, Proc. of the Lepton Photon Symposium,
                Hamburg, July 1997 (to be published).

\bibitem{KARM}  The KARMEN Collaboration, K. Eitel {\it et al.}
                Proc. of the 32nd Rencontre de Moriond, 
                Electroweak Interactions and Unified
                Theories, Les Arcs, March 1997 (to be published),
                hep-ex/9706023
\bibitem{CF}    C. Cardall and G. Fuller, Phys. Rev. {\bf D53},
                4421 (1996).
\bibitem{AP}    A. Acker and S. Pakvasa, Phys. Lett. {\bf B397},
                209 (1997).
\bibitem{DH}    S. Dimopoulos and L. Hall, Phys. Lett. {\bf B207}, 210
                (1987).
\bibitem{GRT}   R. M. Godbole, P. Roy and X. Tata, Nucl. Phys.
                {\bf B401}, 67 (1993).
\bibitem{BC}    G. Bhattacharya and D. Choudhury, Mod. Phys. Lett.
                {\bf A10}, 1699 (1995).
\bibitem{Daw}   S. Dawson, Nucl. Phys. {\bf B261}, 297 (1985).
\bibitem{aleph} ALEPH Collaboration, D.~Buskulic {\it et al.}
                Phys. Lett. {\bf B384}, 461 (1996).
\bibitem{R1}    D. P. Roy, Phys Lett. {\bf B196}, 395 (1987);
                H. Baer, C. Kao and X. Tata, Phys. Rev. {\bf D51}, 2180 (1995);
                M. Guchait and D. P. Roy, Phys. Rev. {\bf D52}, 133 (1995);
                D. Choudhury and S. Raychaudhuri, Phys. Rev. {\bf D56}, 1778
                (1997).
\bibitem{R2} H. Driener, M. Guchait and D. P. Roy, Phys. Rev.  {\bf
                D49}, 3270 (1994); H. Baer, C. Chen and X. Tata,
                Phys. Rev. {\bf D55}, 1466 (1997).
\bibitem{pdg}   R.~M.~Barnett {\it et al.} Phys. Rev. {\bf D54-I}, 1 (1996).
\bibitem{fn1}                  
The reader may worry that imposing symmetries on fields in the
mass basis will restrict
inter-generational mixing. We will see, however, that
the symmetries we impose are approximate in that
they are explicitly broken by some quark Yukawa interactions.
Because of these explicit 
breaking terms, the Yukawa couplings can be exactly the same as in the
Standard Model.
                         

                
                
               
\end{thebibliography}
\end{document}